\documentclass[aps,twocolumn,showpacs,superscriptaddress,prl,10pt]{revtex4-1}
\usepackage{graphicx,braket,amsmath,amsfonts,hyperref,color}

\hypersetup{colorlinks=true,breaklinks=true}

\begin{document}

\title{Low rank representations for quantum simulation of electronic structure}

\date{\today}
\author{Mario Motta} 
\email[Corresponding author: ]{mariomotta31416@gmail.com}
\affiliation{Division of Chemistry and Chemical Engineering, California Institute of Technology, Pasadena, CA 91125}
\author{Erika Ye}
\email[Corresponding author: ]{erikaye@caltech.edu}
\affiliation{Division of Engineering and Applied Sciences, California Institute of Technology, Pasadena, CA 91125}
\author{Jarrod R. McClean}
\affiliation{Google Inc., Venice, CA 90291}
\author{Zhendong Li}
\affiliation{Division of Chemistry and Chemical Engineering, California Institute of Technology, Pasadena, CA 91125}
\author{Austin J. Minnich}
\affiliation{Division of Engineering and Applied Sciences, California Institute of Technology, Pasadena, CA 91125}
\author{Ryan Babbush}
\email[Corresponding author: ]{babbush@google.com}
\affiliation{Google Inc., Venice, CA 90291}
\author{Garnet Kin-Lic Chan}
\email[Corresponding author: ]{gkc1000@gmail.com}
\affiliation{Division of Chemistry and Chemical Engineering, California Institute of Technology, Pasadena, CA 91125}

\begin{abstract}
The quantum simulation of quantum chemistry is a promising application of quantum computers. However, for $N$ 
molecular orbitals, the $\mathcal{O}(N^4)$ gate complexity of performing Hamiltonian and unitary Coupled Cluster 
Trotter steps makes simulation based on such primitives challenging.
We substantially reduce the gate complexity of such primitives through a two-step low-rank factorization of the 
Hamiltonian and cluster operator, accompanied
by truncation of small terms. Using truncations that incur errors below chemical accuracy, we are able to perform 
Trotter steps of the arbitrary basis electronic structure Hamiltonian with $\mathcal{O}(N^3)$ gate complexity in small 
simulations, which reduces to $\mathcal{O}(N^2 \log N)$ gate complexity in the asymptotic regime, while our unitary 
Coupled Cluster Trotter step has $\mathcal{O}(N^3)$ gate complexity as a function of increasing basis size for a 
given molecule.
In the case of the Hamiltonian Trotter step, these circuits have $\mathcal{O}(N^2)$ depth on a linearly connected
array, an improvement over the $\mathcal{O}(N^3)$ scaling assuming no truncation.
As a practical example, we show that a chemically accurate Hamiltonian Trotter step 
for a 50 qubit molecular simulation can be carried out in the molecular orbital basis with as few as 4,000 layers of 
parallel nearest-neighbor two-qubit gates, consisting of fewer than $10^5$ non-Clifford rotations. 
We also apply our algorithm to iron-sulfur clusters relevant for elucidating the mode 
of action of metalloenzymes.
\end{abstract}

\maketitle

The electronic structure (ES) problem, namely, solving for the ground- or low-lying eigenstates of the Schr\"{o}dinger 
equation for atoms, molecules and materials, is an important problem in theoretical chemistry and physics. There are 
several approaches to solving this problem on a quantum computer, including projecting approximate solutions to 
eigenstates using phase estimation \cite{Kitaev1995,Abrams1999,Aspuru-Guzik2005}, directly preparing eigenstates 
using the adiabatic algorithm \cite{Farhi2001,Wu2002,BabbushAQChem}, or using quantum variational algorithms 
\cite{Peruzzo2013,McClean2015,OMalley2016} to optimize parameterized circuits corresponding to unitary Coupled 
Cluster (uCC) \cite{UCC,yanai2006canonical,Romero2017} or approximate adiabatic state preparation 
\cite{Farhi2014,Wecker2015a}.

Time-evolution, under the Hamiltonian or the uCC cluster operator, is a common component in these algorithms. For
near-term quantum devices (especially with limited connectivity), Trotter-Suzuki based methods for time-evolution are
most compelling since they lack the complex controlled operations required by asymptotically more precise 
methods~\cite{Berry2015,Low2016,Low2018}. 
In order to perform a discrete simulation, the Hamiltonian or cluster operator is first represented in a single-particle 
basis of dimension $N$. 
However, in many bases, including the molecular orbital and active spaces bases common in ES, the Hamiltonian and 
cluster operator  contain $\mathcal{O}(N^4)$ second-quantized terms. This leads to at least $\mathcal{O}(N^4)$ gate 
complexity for a single Trotter step~\cite{Seeley2012,Hastings2015}, a formidable barrier to practical progress. While 
complexity can be reduced
using alternative bases~\cite{BabbushLow,Kivlichan2017}, such representations are not usually as compact as the 
molecular orbital one. Thus, reducing the cost of the Trotter step for general
bases is an important goal, particularly within the context of near-term simulation paradigms.

In this Letter, we introduce a general method to reduce the number of gates 
to implement a Trotter step of the Hamiltonian or uCC cluster operator, that 
is especially beneficial for orbital bases, where these operators contain $\mathcal{O}(N^4)$ terms.
Borrowing from classical simulations,
we employ a low-rank decomposition to reduce the Hamiltonian and cluster operator to pairwise  form~\cite{Martinez_2013,DF1,DF2,mCD1,mCD2,mCD3,mCD4,mCD5,mardirossian2018lowering}.
We choose a nested matrix factorization~\cite{Peng_2017} that has an efficient circuit implementation on a quantum computer
via the swap-network strategy~\cite{Kivlichan2017,Jiang2017},
leading to a Hamiltonian Trotter step with an asymptotic gate complexity scaling as 
$\mathcal{O}(N^2\log N)$ with system size, and $\mathcal{O}(N^3)$ for fixed systems and increasing basis size.
These scalings require only linear nearest-neighbor connectivity.
We give numerical evidence that we can carry out a Hamiltonian Trotter step on a 50 qubit quantum chemical problem 
with as few as 4,000 layers of two-qubit gates on a linear nearest-neighbor architecture, a viable target for implementation 
on near-term quantum devices. Compiled to Clifford gates and single-qubit rotations, this requires fewer than $10^5$ 
non-Clifford rotations, an improvement over past Trotter based methods in a fault-tolerant cost model \cite{Reiher2017}.

We first define the  Hamiltonian $H$ and cluster operator $\tau$.
In  second quantization $H$ is
\begin{equation}
\label{eq:gaussian_ham}
	H
    = \sum_{pq=1}^N h_{pq} a^\dagger_p a_q + 
       \frac{1}{2}\sum_{pqrs=1}^N h_{pqrs} a^\dagger_p a^\dagger_q a_r a_s \equiv h+V
\,\,,
\end{equation}
where $a^\dagger_p$ and $a_p$ are fermionic creation and annihilation operators for spin 
orbital $\phi_p$, and the scalar coefficients $h_{pq}$ and $h_{pqrs}$ are the one- and 
two-electron integrals over the basis functions 
$\phi_p$ (here assumed real).

The uCC cluster operator $\tau=T - T^\dagger$, where $T$ is the standard (non-unitary) 
coupled cluster (CC) operator. For uCCSD (uCC with single and double excitations applied to a single determinant reference), 
\begin{align}
\label{eq:ucc}
\tau & = \sum_{i=1}^{N_o} \sum_{a=N_o+1}^{N} t_{ai} (a^\dagger_a a_i- a^\dagger_i a_a) \notag\\
      & + \frac{1}{4} \sum_{ij=1}^{N_o} \sum_{ab=N_o+1}^{N} t_{abij} 
      					(a^\dagger_a a^\dagger_b a_i a_j
- a^\dagger_i a^\dagger_j a_a a_b) \notag \\
&\equiv \sum_{pq=1}^N t'_{pq} a^\dag_p a_q + \frac{1}{4} \sum_{pqrs=1}^N t'_{pqrs} a^\dagger_p a^\dagger_q a_r a_s \quad ,
\end{align}
where $ij$, $ab$ index the $N_o$ occupied and $N_v$ virtual spin orbitals respectively, and  
$N_o+N_v=N$. %
For the scaling arguments with system size, we assume $N_o, N_v \propto N$, while 
increasing basis size corresponds to increasing $N_v$ only. 
Both $H$ and $\tau$ contain $\mathcal{O}(N^4)$ second quantized terms. Thus, 
for arbitrary Hamiltonian integrals or cluster amplitudes, regardless of the gate 
decomposition or fermion encoding used, implementing the time-evolution Trotter step requires at 
least $\mathcal{O}(N^4)$ gates. 

The integrals and cluster amplitudes that one encounters
in molecular ES applications, however, are not arbitrary, but contain considerable structure.
We now show that this allows us to construct approximate operators $H'$ or $\tau'$, accurate 
to within a desired tolerance $\varepsilon$, that can be implemented with greatly reduced gate counts.
The physical
basis for this result is  the pairwise-nature of the Hamiltonian interactions, arising from the $1/r_{12}$ 
Coulomb kernel in real-space.
More precisely, we will rewrite the two-fermion parts of $H$ and $i\tau$ associated
with the integrals $h_{pqrs}$ and $t^\prime_{pqrs}$ as a double-factorized form
\begin{equation}
\label{eq:Vdiag}
\sum_{pq=1}^N S_{pq} a^\dag_p a_q + \sum_{\ell=1}^{L} \sum_{ij=1}^{\rho_\ell} 
\frac{ \lambda_i^{(\ell)} \lambda_j^{(\ell)} }{2} n^{(\ell)}_i n^{(\ell)}_j
\equiv
S + \sum_{\ell=1}^{L} V^{(\ell)} \, ,
\end{equation}
where, defining $\psi^{(\ell)}_i = \sum_{p=1}^N U_{pi}^{(\ell)} \phi_p$,
\begin{equation}
n^{(\ell)}_i 
= 
\sum_{ps=1}^N U_{pi}^{(\ell)} a_p^\dagger a_s U_{si}^{(\ell)} 
= 
a^\dagger_{\psi^{(\ell)}_i} a^{\phantom{\dagger}}_{\psi^{(\ell)}_i}
\end{equation}
are number operators in a rotated basis. Approximate $H'$ and $\tau'$ with reduced complexity can 
then be obtained by truncating the summations over $L$, $\rho_\ell$. The dependence of the 
error $\varepsilon$
on $L$ and $\rho_\ell$ is discussed further below.

The doubly-decomposed form of $V$ can be obtained using a nested matrix factorization, a type of tensor 
factorization introduced in \cite{Peng_2017}.
We illustrate this for the Hamiltonian operator. First, 
the creation and annihilation operators are reordered,
\begin{equation}
V= \frac{1}{2} \sum_{pqrs=1}^N h_{ps,qr} (a_p^\dagger a_s a_q^\dagger a_r - a_p^\dagger a_r \delta_{qs}) = V' + S \, ,
\end{equation}
and $V'$ is recast into a supermatrix indexed by orbitals $(ps),(qr)$ involving electrons 1,2 respectively. 
Due to the eight-fold symmetry 
$h_{pqrs}=h_{srqp}=h_{pqsr}=h_{qprs}=h_{qpsr}=h_{rsqp}=h_{rspq}=h_{srpq}$
this matrix is real symmetric, thus we can write a matrix decomposition in terms of a rank-three auxiliary
tensor $\mathcal{L}$ such that
\begin{equation}
  V'  = \sum_{\ell=1}^{L} \left( \mathcal{L}^{(\ell)} \right)^2 =
  \sum_{\ell=1}^{L} \sum_{pqrs=1}^N \mathcal{L}_{ps}^{(\ell)} \mathcal{L}_{qr}^{(\ell)} 
a_p^\dagger a_s a_q^\dagger a_r .\label{eq:decomp}
\end{equation}
A simple way to obtain $\mathcal{L}$ is to diagonalize the $V'$ supermatrix, although
other techniques~\cite{DF1,DF2,mCD1,mCD2,mCD3,mCD4,mCD5}, such as Cholesky decomposition (CD), 
are also commonly used; we use the Cholesky decomposition in our numerical simulations below.
The next step is to decompose each auxiliary matrix $\mathcal{L}^{(\ell)}$. 
For the Hamiltonian, this is also real symmetric, thus we
can similarly diagonalize it,
\begin{equation}
\begin{split}
	\label{eq:thc}
	\sum_{ps=1}^{N} \mathcal{L}_{ps}^{(\ell)} a_p^\dagger a_s
      &= \sum_{i=1}^{\rho_\ell} U_{pi}^{(\ell)} \lambda_i^{(\ell)} U_{si}^{(\ell)} a_p^\dagger a_s \quad, 
      \end{split}
\end{equation}
where $\lambda^{(\ell)}$, $U^{(\ell)}$ are the eigenvalues and eigenvectors of $\mathcal{L}^{(\ell)}$. 
Combining the two eigenvalue decompositions yields the double-factorized result, 
Eq.~(\ref{eq:Vdiag}). In the cluster operator, amplitudes $t$  have four-fold mixed symmetry 
and antisymmetry, $t_{abij}=t_{jiba}=-t_{baij}=-t_{abji}$. However, as shown in the Supplemental
Information, we can write $i \tau =  \sum_{\ell\mu} {Y}_{\ell,\mu}^2$, where $Y_{\ell,\mu}$ 
are normal and can be diagonalized giving the same double-factorized form.

The double-factorized decomposition Eq.~(\ref{eq:Vdiag}) provides a simple circuit implementation 
of the Trotter step. For example, for the Hamiltonian Trotter step, we write
\begin{align}
  e^{ i \Delta t H} &= 
  e^{i\Delta t ( h+S)} U^{\dag (1)}\prod_{\ell=1}^{L} e^{i\Delta t V^{(\ell)} } 
  \tilde{U}^{(\ell)} + {\cal O}(\Delta t)^2 \,\, ,
  \end{align}
where $\tilde{U}^{(\ell)}=U^{(\ell-1)}U^{\dag (\ell)}$. Time evolution then corresponds to 
(single-particle) basis rotations with evolution under the single-particle operator $h+S$ 
and pairwise operators $V_\ell$. Note that because $h  + S$ is a one-body operator, 
it can be exactly implemented (with Trotter approximation) using a single-particle basis 
change $U^{(0)}$ followed by a layer of $N$ phase gates.
The single-particle basis changes $U^{(\ell)}$ can be implemented using ${N \choose 2} - {N - \rho_\ell \choose 2}$ 
Givens rotations~\cite{Wecker2015b}.
These rotations can be implemented efficiently using two-qubit gates on a linearly connected
architecture~\cite{Kivlichan2017,Jiang2017}. Taking into account $S_z$ spin symmetry to
implement basis rotations separately for spin-up and spin-down orbitals
gives a count of $2{N/2 \choose 2}$ - $2{(N-\rho_\ell)/2 \choose 2}$ 
with a corresponding circuit depth (on a linear architecture) of  $(N + \rho_\ell)/2$.
Using a fermionic swap network, a Trotter step corresponding to evolution under the pairwise operator $V^{(\ell)}$ 
can be implemented in $\rho_\ell \choose 2$ linear nearest-neighbor two-qubit gates, with a two-qubit gate depth 
of exactly $\rho_\ell$. Summing these terms, counts thus are
$N + \sum_\ell \big[ \binom{N}{2} - \binom{N-\rho_\ell}{2} \big] + \binom{\rho_\ell}{2}$
and
$\binom{N}{2} + \sum_{\ell\mu} \big[ \binom{N}{2} - \binom{N-\rho_{\ell,\mu}}{2} \big] + \binom{\rho_{\ell,\mu}}{2}$ for $H$, $\tau$ respectively.

\onecolumngrid

\begin{figure}[t]
\centering
\includegraphics[width=\textwidth]{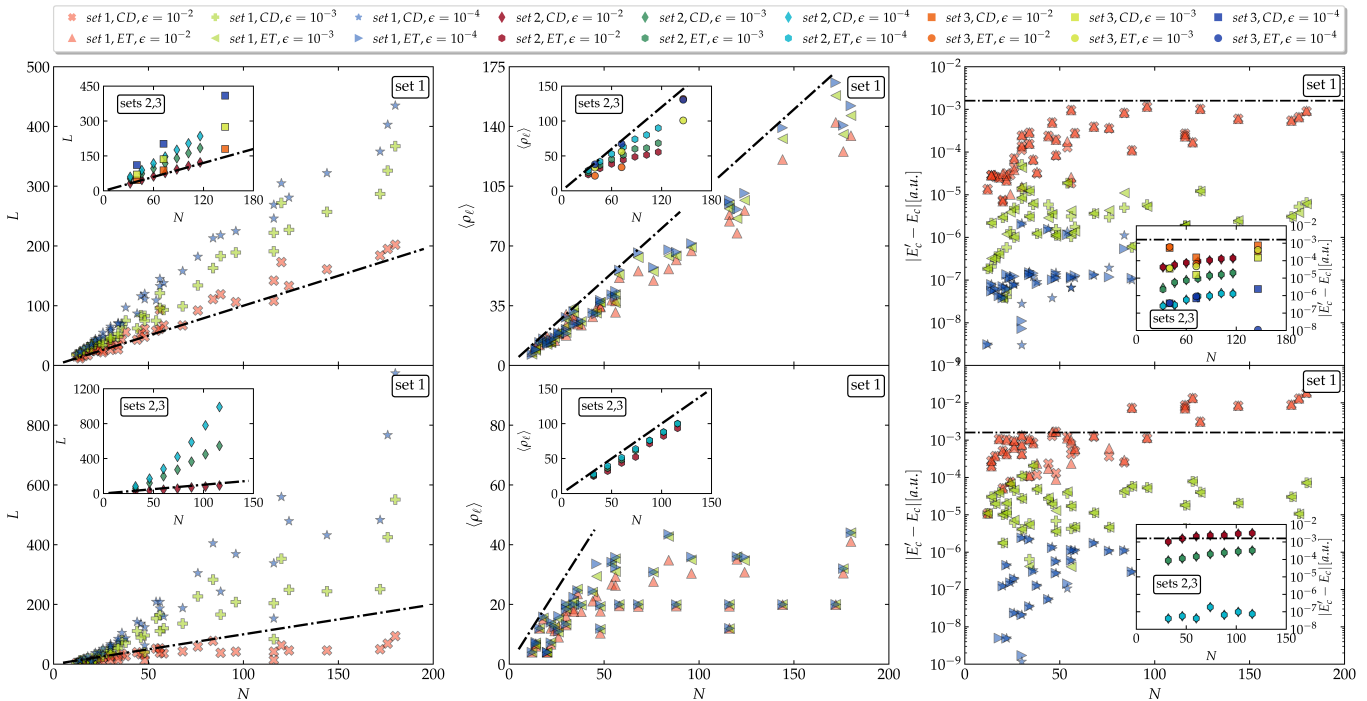}
\caption{Top left: number $L$ of vectors versus basis size $N$. Top middle: average eigenvalue number 
$\langle \rho_\ell \rangle$. 
Top right: error $|E_c^\prime -E_c|$ in the ground-state correlation energy, from the low-rank
approximation of $H$.
Data points in the main figures comprise small molecules with fixed size and 
increasingly large basis (set 1); insets show alkane chains with up to 8 C atoms (set 2)
and iron-sulfur clusters of nitrogenase (set 3).
Lines indicate $N$ (left, middle) and the chemical accuracy (right). Bottom: 
same as the upper panel, for the uCC operator $\tau$, with $\langle \rho_\ell \rangle$ averaged over $\mu$. }
\label{fig:results}
\end{figure}
\twocolumngrid

To realize this algorithm on a near-term device, where the critical cost model is the number of 
two-qubit gates, one can either
implement the gates directly in hardware \cite{Murphy2018}, which requires
$\sum_{\ell=1}^{L} \big[ \frac{N \rho_\ell}{4} + \frac{\rho_\ell^2}{4} - \rho_\ell \big]$ 
gates on a linear nearest neighbor architecture, with circuit
depth $\sum_{\ell=1}^L \frac{N}{2} + \frac{3\rho_\ell}{2}$.
If decomposing into a standard two-qubit gate set (e.g. CZ or CNOT), the gate count would be 
three times the above count.

To realize this algorithm within an error-corrected code such as the surface code \cite{Fowler2012}, 
where the critical cost model is the number of T gates, one can decompose each Givens rotation 
gate in two arbitrary single-qubit rotations and each diagonal pair interaction in one arbitrary 
single-qubit rotation. Thus, the number of single-qubit rotations is 
$\sum_{\ell=1}^{L} \big[ \frac{N \rho_\ell}{2} - 2 \rho_\ell \big]$.
Using standard synthesis techniques the number of T gates is then $2.3 \log(1/\varepsilon)$ times 
this count \cite{Bocharov2015}.

For an exact decomposition of $H$, $L=N^2$ and $\rho_\ell=N$. However, it is well established 
from empirical ES calculations that both $L$ and $\rho_\ell$ can be 
significantly truncated if we approximate $H$ and $\tau$ by $H'$ and $\tau'$ with error $\varepsilon$.
In the case of $L$, we rely on the CD and perform the truncation based on the $L^\infty$ norm, 
i.e. use the smallest $L$ such that 
$\max_{psqr} |h_{ps,qr} - \sum_{\ell=1}^L \mathcal{L}^{(\ell)}_{ps} \mathcal{L}^{(\ell)}_{qr} |< \varepsilon$.
For $\rho_\ell$, we perform an eigenvalue truncation (ET) based on the $L^1$ norm, i.e. 
use the smallest $\rho_\ell$ such that $\sum_{j=\rho_\ell+1}^{N} | \lambda_j^{(\ell)} | < \varepsilon$. 
For this truncation of $H$ it has been shown that, when increasing the molecular size or simulation 
basis, $L \sim \mathcal{O}(N)$, while $\langle \rho_\ell \rangle = \frac{1}{L} \sum_{\ell=1}^L \rho_\ell \sim 
\mathcal{O}(\log N)$ for increasing molecular size~\cite{Peng_2017}. 
For the uCC operator, using antisymmetric amplitudes as in this work yields a different 
scaling of the Cholesky decomposition, where $L \sim \mathcal{O}(N)$ with increasing basis but 
$L \sim \mathcal{O}(N^2)$ with increasing molecular size (albeit with a small coefficient); 
the scaling properties of $\rho_{\ell,\mu}$ have not previously been studied.

In Fig.~\ref{fig:results} we show $L$ and $\langle \rho_\ell \rangle$ for different truncation 
thresholds in: 
(set 1) a variety of molecules that can be represented with a modest number 
of qubits (CH$_4$, H$_2$O, CO$_2$, NH$_3$, H$_2$CO, H$_2$S, F$_2$, BeH$_2$, HCl)
using STO-6G, cc-pVDZ, 6-31G*, cc-pVTZ bases; 
(set 2) alkane chains C$_n$H$_{2n+2}$, 
$n \leq 8$, using the STO-6G basis; 
(set 3) Fe-S clusters ([2Fe-2S], [4Fe-4S], and the 
nitrogenase P$^\text{N}$ cluster) in active spaces with $N=40$, 72, 146 respectively.
Details of calculations are given in the Supplemental Information.
For the uCC operator, we have used the (classically computable) traditional CC amplitudes, 
equal to the uCC amplitudes in the weak-coupling limit.  
For $H$, we clearly see the $L \propto N$ scaling across different truncation thresholds, 
for both increasing system size and basis. For $\tau$, $L \propto N$ with increasing 
basis in a fixed molecule, while $L \propto N^2$ with increasing size (e.g. in alkane chains).
Interestingly, the value of $L$ in the Hamiltonian decomposition is quite similar across 
different molecules for the same number of spin-orbitals (qubits).
In the subsequent ET for the Hamiltonian, $\langle \rho_\ell \rangle$ features 
$\mathcal{O}(\log N)$ scaling for alkanes ($n \geq 5$, represented here with 75-125 qubits).
For the uCC operator, we observe that $\langle \rho_{\ell} \rangle$ scales like 
$\mathcal{O}(N)$ for alkane chains and increasing molecular size, while it is 
approximately constant for increasing basis set size.
The less favourable scaling of $L$, $\langle \rho_{\ell,\mu} \rangle$ with system size 
for the uCC operator, relative to $H$, stems from the antisymmetry properties of the 
amplitudes, which in the current factorization means that $Y_{\ell,\mu}$ do not show 
the same sparsity as $\mathcal{L}^{(\ell)}$.

\begin{figure}[t]
\centering
\includegraphics[width=0.43\textwidth]{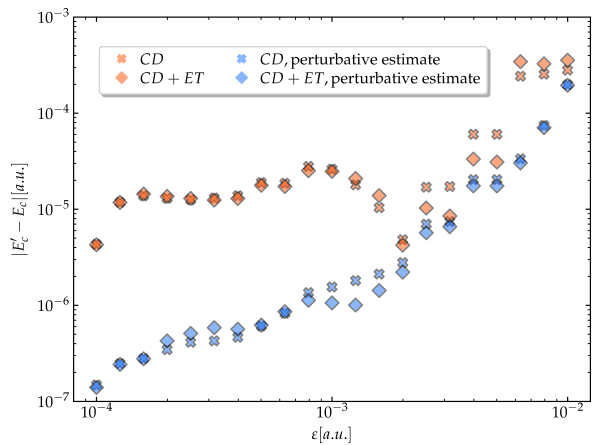}
\caption{Error $|E^\prime_c - E_c|$ as a function of threshold $\varepsilon$, with and without 
perturbative correction (blue, orange points) for CD and CD+ET truncation schemes (crosses, 
diamonds), measured using HF and CC wavefunctions.}
\label{fig:acc}
\end{figure}

The error arising from the truncations leading to $H'$ and $\tau'$ can be understood in terms of two components: 
(i) the error in the operators, and (ii) the error in the states generated by time-evolution with these operators.
It is possible to substantially reduce both errors using quantities that can be computed classically.
We illustrate this for the error in $H'$. 
First, the correlation energy, defined as $E_c = E - E_\text{HF}$,
with $E$ the total energy and $E_\text{HF}$ the Hartree-Fock energy, is usually a much smaller quantity
than the total energy in chemical systems, and is affected by much smaller truncation errors, mainly
due to cancellation or errors between exact and mean-field truncations. Thus, 
using the classically computed mean-field energy of $H'$, we can obtain 
the truncated correlation energy as $E_c' = E' - E_\text{HF}'$.
Second, we can estimate the remaining error in $E^\prime_c$ from first-order perturbation theory as 
$\langle \psi | H-H' | \psi \rangle $ using a classical approximation to $\psi$; if the classical $\psi$ is 
accurate, the corrected $E^\prime_c$ is then accurate to $\mathcal{O}(\varepsilon^2)$. 
In Fig.~\ref{fig:acc} we plot $|E^\prime_c - E_c|$ for $\mathrm{H_2O}$ at the cc-pVDZ level.
Adding the perturbative correction from the classical CC ground-state reduces the error by about an 
order of magnitude, such that even an aggressive truncation threshold of $\varepsilon = 10^{-2}$ 
$\mbox{a.u.}$
yields the total correlation energy within the standard chemical accuracy of $1.6 \times 10^{-3}$ 
$\mbox{a.u.}$
For the $\tau^\prime$ truncation, one could include a similar error correction for the correlation 
energy derived from approximate cluster amplitudes, although we do not do so here.

\begin{figure}[b]
\centering
\includegraphics[width=0.43\textwidth]{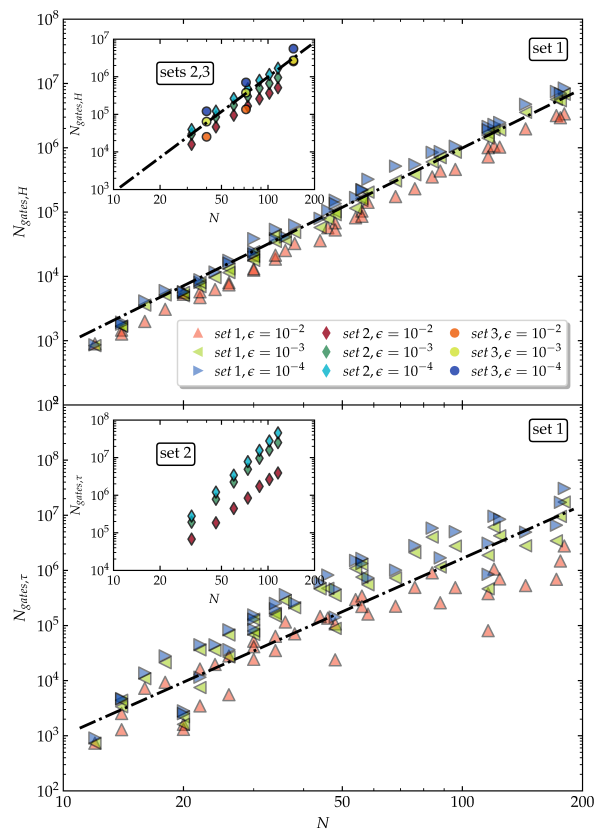}
\caption{Gate counts per Trotter step of the Hamiltonian (top) and uCC operator (bottom), 
for $\varepsilon = 10^{-2}$, $10^{-3}$ and $10^{-4}$ (red, green, blue).
Black lines indicate power-law fits, with optimal exponents 3.06(3) and 3.21(9) (top, bottom).}
\label{fig:gc_h}
\end{figure}

In Fig.~\ref{fig:gc_h} we show the total gate counts needed to carry out a Trotter step of $H'$ and $\tau'$ 
with different truncation thresholds.
Using the scaling estimates obtained above for $L, \rho_\ell$ in the gate count expression,
we expect the Hamiltonian Trotter step to have a gate count $N_{gates} \sim \mathcal{O}(N^2 \log N)$ 
for increasing molecular size,
and $\mathcal{O}(N^3)$ for fixed molecular size and increasing basis size, and the uCC 
Trotter step to show $N_{gates} \sim \mathcal{O}(N^4)$ for increasing molecular size
and $\mathcal{O}(N^3)$ with increasing basis size.
This scalings are confirmed by the gate counts in Fig.~\ref{fig:gc_h}.
As seen, the crossover between $N^3$ and $N^2 \log N$ behavior of the Hamiltonian Trotter 
gate cost, for alkanes, occur at larger $N$ than one would expect from the $\langle \rho_\ell \rangle$ data 
alone from Fig.~\ref{fig:results}, due to tails in the distribution of $\rho_\ell$. 

The threshold for classical-quantum crossover in general purpose computation
is usually considered to occur at 50 qubits. For near-term devices, the number of layers of gates on a parallel 
architecture with restricted connectivity is often considered a good cost model. Using the circuit depth estimate 
$\sum_\ell (N+\rho_\ell)$, we see that we can carry out a single Hamiltonian Trotter step on a 
system with 50 qubits with as few as 4,000 layers of parallel gates on a linear architecture.
Within cost models appropriate for error-correction, the most relevant cost metric is the number of T gates 
\cite{Childs2017,Reiher2017,BabbushSpectra}.
For our algorithms, T gates enter through single-qubit rotations and thus, the number of non-Clifford single-qubit 
rotations is an important metric. Based on the gate count estimate for basis changes, the number of 
non-Clifford rotations required for our Trotter steps is roughly 100,000 and the number of T gates per 
rotation is approximately 20-50 times this number.
  
In summary, we have introduced a nested decomposition of the Hamiltonian and uCC operators,
leading to substantially reduced gate complexity for the Trotter step both in realistic
molecular simulations with under 100 qubits, and in the asymptotic regime.
The discussed decomposition is by no means the only one possible and, for the uCC operator, it is 
non-optimal, as more efficient decompositions for antisymmetric quantities exist \cite{Shenvi2013}. 
Future work to better understand the interplay between classical tensor decompositions and the 
components of quantum algorithms thus presents an exciting possibility for further improvements 
in practical quantum simulation algorithms.

{\em{Acknowledgments.}}

RB, GKC and MM contributed to the conception of the project; 
MM, JRM, ZL, RB and GKC contributed to the theoretical analysis; MM, ZL, EY and GKC contributed
to the electronic structure calculations; MM, JRM, EY, RB and GKC contributed to the gate counts 
analysis. All authors contributed to the drafting of the paper.

MM, ZL, and GKC (theoretical analysis, electronic structure calculations, drafting of the paper) were
supported by NSF grant number 1839204. EY (gate counts, electronic structure calculations) was 
supported by a Google graduate fellowship and a Google award to GKC. AM (drafting of the paper) 
was supported by NSF grant CBET CAREER number 1254213.

MM gratefully acknowledges Shiwei Zhang and James Shee for valuable interaction.

%\bibliographystyle{apsrev4-1}
%\bibliography{art}


\pagebreak
\newpage

\section{Supplemental Information}

\subsection{Low-rank factorization of uCC amplitudes}
The uCC operator has the expression $\tau = t-t^\dagger$, where
\begin{equation}
t = \frac{1}{4}
\sum_{ij=1}^{N_o} \sum_{ab=N_o+1}^N 
t_{ab ij} 
\,
a^\dagger_a a^{\phantom{\dagger}}_j a^\dagger_b a^{\phantom{\dagger}}_i 
\quad ,
\label{eq:dec0}
\end{equation}
and the tensor $t_{abij}$ has the symmetry properties 
$t_{abij} = - t_{baij} = - t_{abji}$, readily implying $t_{abij} = t_{baji}$.
The first step to represent $\tau$ as a linear combination of squares of 
normal operators, similarly to the first decomposition of the electron 
repulsion integral, is to represent $t$ as
\begin{equation}
t = 
\sum_{j<i} \, \sum_{a<b} 
t_{ab ij} 
\,
a^\dagger_a a^{\phantom{\dagger}}_j a^\dagger_b a^{\phantom{\dagger}}_i 
\quad .
\end{equation}
This equation can be written exactly as
\begin{equation}
t = 
\sum_{psqr=1}^N T_{ps,qr}
\,
a^\dagger_p a^{\phantom{\dagger}}_s a^\dagger_q a^{\phantom{\dagger}}_r 
\quad ,
\end{equation}
having introduced a highly sparse tensor $T$, 
\begin{equation}
T_{ps,qr} = 
\left\{
\begin{array}{ll}
t_{pqrs} & \mbox{if $p<q$, $p,q > N_o$, $s<r$, $r,s \leq N_o$} \\
0 & \mbox{otherwise} \\
\end{array}
\right.
.
\end{equation}
Performing a singular value decomposition of $T$ \cite{LiJCP144} leads to
$T_{ps,qr} = \sum_{\ell=1}^L \sigma_\ell \, U^{(\ell)}_{ps} V^{(\ell)}_{qr}$ and
\begin{equation}
\tau = \sum_{\ell=1}^L \left( U_\ell V_\ell - V^\dagger_ \ell U^\dagger_ \ell \right) \quad ,
\label{eq:dec2}
\end{equation}
where $U_\ell = \sum_{ps} \sqrt{\sigma_\ell} \, U^{(\ell)}_{ps} a^\dagger_p a_s$
and similarly for $V_\ell$. 

As stated and illustrated in the main text, the number of retained
eigenvalues scales like $\mathcal{O}(N^2)$ for increasing system size. This
less favorable scaling for the uCC operator, relative to $H$, stems from the 
antisymmetry properties of the amplitudes. This can be understood observing
that, at second-order perturbation theory in the electron-electron interaction,
uCC amplitudes are $t_{abij} \propto \frac{\langle ji || ab \rangle}{\epsilon_a - 
\epsilon_j + \epsilon_b - \epsilon_i}$, where $\epsilon_p$ are the Hartree-Fock 
eigenvalues and $\langle ji || ab \rangle = h_{abij} -h_{abji}$ the antisymmetrized 
electron repulsion integral. Antisymmetrization of the electron repulsion integral
prevents from casting $t$ onto a supermatrix where indexed by orbitals 
pertaining to electrons 1 and 2 separately.

Note that, since $U^{(\ell)}_{ps} \neq 0$ unless $p > N_o$ and $s \leq N_o$, 
$[U_\ell , V_\ell ] = 0$. Therefore, we can write
\begin{equation}
\begin{split}
\tau = \frac{1}{4} \sum_{\ell=1}^L 
&\left( U_\ell + V_\ell \right)^2 - \left( U_\ell^\dagger + V_\ell^\dagger \right)^2 \\
- &\left( U_\ell - V_\ell \right)^2 + \left(U_\ell^\dagger - V_\ell^\dagger \right)^2
\quad ,
\end{split}
\label{eq:dec2}
\end{equation}
Equation \eqref{eq:dec2} expresses $\tau$ as a sum of products,
of the form $X^2 - \left(X^\dagger \right)^2$, where $X$ is not a normal operator. 
The identity
\begin{equation}
\begin{split}
X^2 -\big( X^\dagger \big)^2 
&= i \left( \frac{1-i}{2} \big(X + i X^\dagger \big) \right)^2 \\
&- i 
\left( \frac{1+i}{2} \big(X - i X^\dagger \big) \right)^2 
\end{split}
\label{eq:dec3}
\end{equation}
of easy verification, leads to squares of normal operators. 
Applied to \eqref{eq:dec2},  \eqref{eq:dec3} leads immediately to
\begin{equation}
i \tau = \sum_{\ell=1}^L \sum_{\mu=1}^4 Y_{\ell,\mu}^2 
\label{eq:dec4}
\end{equation}
where, for each $\ell$, the four $Y_{\ell,\mu}$ operators 
\begin{equation}
\begin{split}
Y_{\ell,0} &= \phantom{i} \frac{1+i}{32} \Big( (U_\ell + V_\ell) - i (U_\ell + V_\ell)^\dagger \Big) \\
Y_{\ell,1} &= i \frac{1-i}{32} \Big( (U_\ell + V_\ell) - i (U_\ell + V_\ell)^\dagger \Big) \\
Y_{\ell,2} &= \phantom{i} \frac{1+i}{32} \Big( (U_\ell - V_\ell) - i (U_\ell - V_\ell)^\dagger \Big) \\
Y_{\ell,3} &= i \frac{1-i}{32} \Big( (U_\ell - V_\ell) - i (U_\ell - V_\ell)^\dagger \Big) \\
\end{split}
\end{equation}
are normal, 
and proportional to $(U_\ell + V_\ell) \pm (U_\ell + V_\ell)^\dagger$ or 
$(U_\ell - V_\ell) \pm (U_\ell - V_\ell)^\dagger$ depending on the value
of $\mu$.
Eq \eqref{eq:dec4} expresses $i \tau$ as $i$ times a linear combination of 
squares of normal operators, the form used in the main text.

Eigenvalues such that $|\sigma_\ell| < \varepsilon$ can be discarded in the
spirit of the CD for the Hamiltonian. 
For the uCC operator, the subsequent ET procedure is carried out as follows. 
In the basis of Hartree-Fock orbitals, the operators $Y_{\ell,\mu}$ are
described by the matrices
\begin{equation}
\left( Y_{\ell,\mu} \right)_{pr} = 
\left(
\begin{array}{c|c}
0 & T_{\ell,\mu} \\
\hline
T^\dagger_{\ell,\mu} & 0 \\
\end{array}
\right)
\quad.
\end{equation}
where $T_{\ell,\mu}$ is proportional to $U_\ell + V_\ell$ or $U_\ell + V_\ell$

Introducing the SVD of 
$T_{\ell,\mu} = A \, \mbox{diag}( \rho_{\ell,\mu} ) B^\dagger$,
it easy to verify that the only non-zero eigenvalues of $Y_{\ell,\mu}$
have the form
\begin{equation}
\begin{split}
v^i_{\ell,\mu} &= \frac{1}{\sqrt{2}} 
\left(
\begin{array}{c}
a_i \\
\hline
b_i \\
\end{array}
\right)
\quad\quad,\quad
Y_{\ell,\mu} v^i_{\ell,\mu}
= \rho^i_{\ell,\mu} \, v^i_{\ell,\mu}
\quad ,      \\
w^i_{\ell,\mu} &= \frac{1}{\sqrt{2}} 
\left(
\begin{array}{c}
\phantom{-} a_i \\
\hline
- b_i \\
\end{array}
\right)
\quad,\quad
Y_{\ell,\mu} w^i_{\ell,\mu}
= - \rho^i_{\ell,\mu} \, w^i_{\ell,\mu}
\quad . \\
\end{split}
\end{equation}
where $a_i$, $b_i$ are the $i$-th columns of $A$, $B$ respectively.
The ET is performed truncating the singular values $\rho^i_{\ell,\mu}$ of $T_{\ell,\mu}$ 
and retaining the largest $\rho_{\ell,\mu}$ of them.

\subsection{Partial Basis Rotation}
Suppose we have a single particle basis rotation given by $U^{(\ell)}$, such that 
\begin{equation}
\tilde{a}_p^\dagger = \sum_{q=1}^N U_{pq}^{(\ell)} a_q^\dagger \qquad 
\tilde{a}_p = \sum_{q=1}^N \left( U_{pq}^{(\ell)}\right)^* a_q
\end{equation}
The Thouless theorem \cite{Thouless1960} provides a means for implementing basis rotations 
on quantum computers. In essence, one applies a series of rotations that act on 
two rows ($p,q$) of $U^{(\ell)}$ at a time. The series of rotations is determined by performing a QR 
decomposition of $U^{(\ell)}$ using Givens rotations $r_{pq}(\theta_{pq})$.  One only needs to 
perform rotations on the $N\choose 2$ elements below the diagonal of $U^{(\ell)}$ and, when 
done in the correct order, these can be performed in linear depth on a device with linear connectivity
\cite{Kivlichan2017}.

In this Appendix, we show that one can perform an approximation basis transformation using on 
the order of $\rho_\ell N$ rotations, where $\rho_\ell \leq N$ and is the number of eigenvalues 
retained after making a low-rank approximation.

Consider the eigenvalue equation used in order to obtain $U^{(\ell)}$,
\begin{equation}
\sum_{q=1}^N \mathcal{L}^{(\ell)}_{pq} U^{(\ell)}_{qi} = \lambda^{(\ell)}_i U^{(\ell)}_{pi} \quad,
\end{equation}
where $\mathcal{L}^{(\ell)}$ is the $\ell$-th Cholesky vector, reshaped into a square matrix of order $N$, 
and  $\lambda^{(\ell)}$ is the diagonal matrix of the corresponding eigenvalues. We choose $U^{(\ell)}$ 
such that the numbers $\lambda^{(\ell)}$ are in decreasing order of magnitude.

In the main text, we perform a low-rank approximation of $\mathcal{L}^{(\ell)}$ by considering only the 
$\rho_\ell$ largest eigenvalues or, equivalently, the eigenvectors associated with the first $\rho_\ell$ 
columns of $U^{(\ell)}$. This effectively reduces the sizes of the matrices involved, 
and the eigenvalue equation becomes
\begin{equation}
\Big( \bar{U}^{(\ell)} \Big)^\dagger L^{(\ell)} \bar{U}^{(\ell)} = \bar{\lambda}^{(\ell)}_i \quad,
\end{equation}
where $\bar{U}$ and $\bar{\lambda}$ are matrices comprising the first $\rho_\ell$ columns of $U$ and 
$\lambda$ respectively.  Now, we only need to perform a QR decomposition of $\bar{U}^{(\ell)}$, after 
which the top $\rho_\ell \times \rho_\ell$ block is diagonal and the lower $(N-\rho_\ell)$ rows are zero. 
Since $\bar{U}^{(\ell)}$ is only an $N \times \rho_\ell$ matrix, fewer Givens rotations are needed than
in the general case. 
Specifically, the QR decomposition at worst requires 
$N\rho_\ell - \rho_\ell(\rho_\ell+1)/2$ rotations, as that is the number of terms below the diagonal,
coinciding with the estimate $\binom{N}{2} - \binom{N-\rho_\ell}{2}$ given in the main text.

In the quantum algorithm proposed in the main text, one rotates from the basis of one Cholesky vector to another.
This rotation, explicitly given by $U^{(\ell+1)} U^{(\ell)}$, can be reduced into a single unitary operation, 
and the cost of approximately implementing the basis rotation as described here is determined by $\rho_{\ell+1}$.

\subsection{Details of calculations}

In this Appendix, we provide further details about the calculations yielding the data showed in the 
main text, focussing on each of the three studied sets (set 1, set 2, set 3). 

{\em{Set 1 -- }} comprises 9 small molecules (namely CH$_4$, H$_2$O, CO$_2$, 
NH$_3$, H$_2$CO, H$_2$S, F$_2$, BeH$_2$, HCl), studied at experimental equilibrium geometries
from \cite{NIST}.
Molecules in this "set 1" have been studied with restricted Hartree Fock (RHF) and restricted
classical coupled cluster with single and double excitations (RCCSD) on top of the RHF state.
Matrix elements of the Hamiltonian and classical RCCSD amplitudes have been computed with
the PySCF software \cite{PySCF}, using the STO-6G, 6-31G*, cc-pVDZ, cc-pVTZ bases.

{\em{Set 2 -- }} comprises alkane chains (namely ethane, propane, butane, pentane, hexane,
heptane and octane, all described by the chemical formula $C_n H_{2n+2}$ with $n=2 \dots 8$),
studied at experimental equilibrium geometries from \cite{NIST}.
Molecules in this "set 2" have been studied with RHF, RCCSD methods.
Matrix elements of the Hamiltonian and classical RCCSD amplitudes have been computed with
the PySCF software \cite{PySCF}, using the STO-6G basis.

\begin{figure}[h!]
\centering
\begin{tabular}{ccc}
\includegraphics[width=0.95\columnwidth]{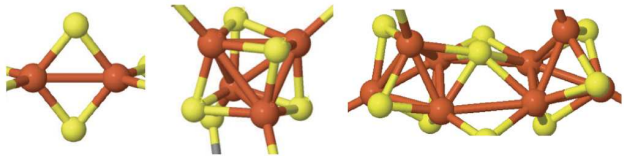} \\
\\
{\scriptsize (a) [2Fe-2S] (30e,20o) \,\, \scriptsize (b) [4Fe-4S] (54e,36o) \,\, \scriptsize (c) [8Fe-7S] (114e,73o) } \\
\end{tabular}
\caption{Iron-sulfur clusters used in the present work, and their active spaces (specified by
numbers of active electrons and orbitals). } \label{fes}
\end{figure}

{\em{Set 3 -- }} comprises Fe-S clusters [2Fe-2S] [2Fe(II)] and [4Fe-4S] [2Fe(III),2Fe(II)], 
and the P$^\mathrm{N}$ cluster [8Fe-7S] [8Fe(II)]) of nitrogenase.

The active orbitals of [2Fe-2S] and [4Fe-4S] complexes were prepared by a split 
localization of the converged molecular orbitals at the level of BP86/TZP-DKH, 
while those of the [8Fe-7S] were prepared at the level of BP86/def2-SVP. The active 
space for each complex was composed of Fe 3$d$ and S 3$p$ of the core part and $\sigma$-bonds 
with ligands, which is the minimal chemically meaningful active space. The structure of the 
iron-sulfur core and the numbers of active orbitals and electrons for each complex are 
summarized in Figure \ref{fes}, and the matrix elements $h_{pq}$, $h_{pqrs}$ are made available
in a compressed archive form (FeS\_integrals.tar).

Molecules in this "set 3" were treated with density-matrix renormalization group (DMRG) 
\cite{DMRG,DMRG2}, using the PySCF software.
The DMRG calculations were performed for the $S=0$ states, which are the experimentally identified 
ground states, with bond dimensions 8000, 4000, and 2000 for [2Fe-2S], [4Fe-4S], and 
[8Fe-7S]. Note that the active space employed in the present work for the P-cluster is larger than the
active space previously used to treat the FeMoco cluster of nitrogenase, having the same number of
transition metal atoms \cite{Reiher2017}.

Broken-symmetry unrestricted Hartree-Fock (UHF) ($M_S=0$) calculations were carried 
out for [2Fe-2S] and [4Fe-4S]. For [8Fe-7S], due to convergence issues,
high-spin UHF calculations were used instead.

\end{document}